# BPM for the masses: empowering participants of Cognitive Business Processes

Aleksander Slominski and Vinod Muthusamy

IBM T.J. Watson Research Center, Yorktown Heights NY, 10598, USA
`{aslom,vmuthus}@us.ibm.com`

**Abstract.** Authoring, developing, monitoring, and analyzing business processes has requires both domain and IT expertise since Business Process Management tools and practices have focused on enterprise applications and not end users. There are trends, however, that can greatly lower the bar for users to author and analyze their own processes. One emerging trend is the attention on blockchains as a shared ledger for parties collaborating on a process. Transaction logs recorded in a standard schema and stored in the open significantly reduces the effort to monitor and apply advanced process analytics. A second trend is the rapid maturity of machine learning algorithms, in particular deep learning models, and their increasing use in enterprise applications. These cognitive technologies can be used to generate views and processes customized for an end user so they can modify them and incorporate best practices learned from other users' processes.

**Keywords:** BPM, cognitive computing, blockchain, privacy, machine learning

## 1. End Users and Business Process Management

Business Process Management (BPM) tools and techniques address the life cycle of developing, executing, and managing business processes [1]. By capturing process execution traces inside a BPM runtime it was possible to get better visibility into process state and address problems faster. Moreover, gathered data could be used to improve processes and create better outcomes for process participants. However, the greater visibility into process execution was not available for participants that do not have direct access to the BPM runtime. Instead, they get what is made visible through UI interactions but rarely have any access to process state or historical data. When there are multiple participants in a business process from different organizations, each participant may have their own BPM system and have limited or no visibility into other systems.

A traditional challenge with constructing a business process is the specification of standard schemas that multiple parties can agree on, sometimes referred as the Process Reuse challenge [1]. We argue that the use of blockchain forces parties collaborating in a business process to develop, and more importantly, agree on the schema. This, in turn, offers end-to-end visibility into the process and the ability for new participants to devise ad-hoc auxiliary business processes without permission from a central coordinator. Our vision is that this could be as significant as the introduction of spreadsheets for PCs in creating a programming model that allows casual users to create processes and analyze data without becoming computer science or data science experts.

## 2. Use case: health care

Consider a health care scenario where there are multiple parties involved in the care of a patient, including the hospital, physicians, nurses, lab technicians, insurance agency, government auditors, the patient, and their family members. A business process that manages a patient's visit needs to coordinate activities among all these participants without violating privacy policies.

It is extremely unlikely that one business process will capture all the rules and requirements among these parties. What is more practical is for subsets of parties to devise processes that help them get their work done. For example, a hospital, insurance agency, and auditor may develop a process to manage insurance policy requests and payment invoices.

The power of using a shared blockchain to maintain process state using a public schema is that it allows new processes to be added without having to navigate a centralized governance process.

**Fig. 1.** This screenshot is an illustration of how a process end user can specify event triggers and generate relevant notifications that may send alerts to their mobile device.

As data is added to the overall end-to-end process blockchain, multiple participants with access to the data can delegate it to pieces of code that run automatically. The code can be a smart contract that does something when conditions are met in the blockchain. For example, the trigger can be the event that Patient A enters hospital area, as in Fig. 1. Alternatively, end users can directly consume updates to the blockchain by receiving events and having apps on their mobile device perform a custom action. Another option is for events to be passed to other services that a user has authorized to access some data from their part of blockchain.

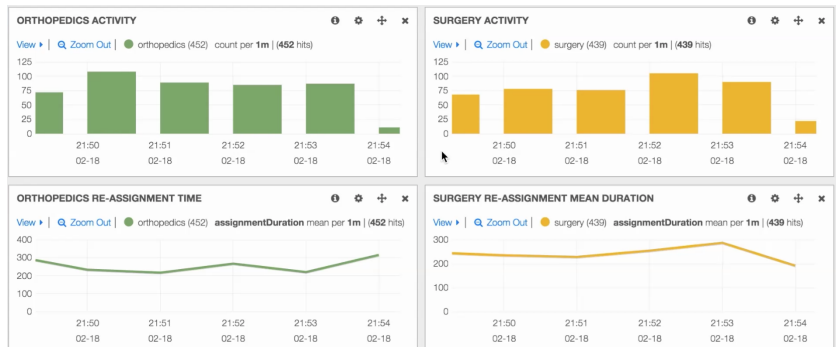

**Fig. 2.** This screenshot shows how blockchain data could be leveraged to build analytics solutions that cross traditional boundaries and silos. In this hypothetical example, data from two separate hospital divisions can be put into one analytics dashboard even though each division may run its own BPM system but they share state in blockchain.

Storing even a small amount of relevant data in a blockchain allows any participant to build end-to-end dashboards and analytics capabilities. For example, in this health care use case, the above dashboard compares key metrics in two departments using data that could be stored in a blockchain (Fig. 2).

We want to emphasize the importance of the data being in a shared blockchain that any authorized user can access. This supports innovative applications that make use of the data for discovering inefficiencies, enforcing smart contracts, auditing for compliance with regulations, or building reactive user experiences.

## 3. Privacy in Cognitive BPM

Privacy controls are a major challenge in many enterprise applications. Fortunately, it possible to not only to store process state in a blockchain, but also related permissions and have the sensitive data encrypted and made accessible only to authorized parties. Encrypting data, however, makes it difficult to perform analytics, and hence decreases one of main advantages of sharing state.

Consider a health case scenario that exemplifies some of the challenges: patients, health providers, insurance agencies, and government auditors all need different views of a patient's sensitive information.

There are well-known security technologies, including differential privacy [2], homomorphic encryption, role-based access control, and pseudo-anonymity, that can be used to address some of the important privacy issues. The same issues exist in cloud computing in general [3]. The use of these and other techniques to blockchain applications is still an open problem.

## 4. Machine learning to guide process development

Developing and refining a business process can be difficult and time consuming. It requires domain expertise to understand how to optimize the process logic, as well as technical proficiency to formally encode the process logic. We cannot assume that end users will be skilled in either of these dimensions, but there are machine learning techniques that can be used to lower the barrier.

One technique takes advantage of the fact that we can observe in the blockchain the end-to-end behavior of processes. We can, therefore, extract individual process traces, find recurring patterns, and generate business rules or process models that capture this behavior. Existing process mining techniques can be applied here [4,5]. There is still, however, an open problem on how to present these process templates at a level of abstraction that is both easy to understand for the user, and gives them the ability to customize it.

The above technique assumes that the process models themselves are not shared, but only their effects. In cases where users are willing to share their custom processes and rules, we can apply different techniques to find the ones that are performing well (according to some business metric) and suggest these to other users. Clustering and recommendation algorithms, among others, in the process analytics domain can be used here [6].

Consider an example of a coach helping a patient reach a fitness goal, such as recovering from a heart surgery. The coach develops a customized plan that includes defines concrete tasks and reminders related to their diet, physical therapy, medication and lab visits. This plan is essentially a set of business rules and processes that requires collaboration among the coach, patient, and other health providers. The participants may further refine the process as needed. While each patient will need a customized process, there are probably patterns that emerge on the kinds of plans that work well with certain classes of patients. Machine learning algorithms can help find these best practices and suggest them to the right users.

We believe that the visibility into the behavior of blockchain-based applications offers a rich, and unprecedented, amount of data that can be mined to help simplify and guide end users who want to develop a business process.

## 5. Conclusions

Using blockchain to orchestrate cognitive BPM processes not only make them easier for cognitive analytics but make them more accessible to process participants. By developing an ecosystem of apps and services that can get access to cognitive BPM state in blockchain, users will have a marketplace of solutions and new types of apps that are much better at understanding users and what they are doing by understanding how they participate in cognitive processes.

**References**

1. Aalst W.M.P.: Business Process Management: A Comprehensive Survey. In: ISRN Software Engineering (2013).
2. Narayanan A., Shmatikov V.: Robust De-anonymization of Large Sparse Datasets. In: Proceedings of the IEEE Symposium on Security and Privacy (2008).
3. Chen D, Zhao H.: Data security and privacy protection issues in cloud computing. In: International Conference on Computer Science and Electronics Engineering (2012).
4. Muthusamy, V., Slominski A., Ishakian V., Khalaf R., Reason, J., Rozsnyai, S.: Lessons learned using a process mining approach to analyze events from distributed applications. In: Proceedings of the 10th ACM International Conference on Distributed and Event-based Systems (2016).
5. Evermann, J.-R., Fettke, P.: Predicting Process Behaviour Using Deep Learning. In: Decision Support Systems (2017).
6. Ehrig, M., Koschmider, A., Oberweis, A.: Measuring similarity between semantic business process models. In: Proceedings of the fourth Asia-Pacific conference on Conceptual modelling - Volume 67 (2007).